\documentclass[12pt]{article}
\usepackage[dvips]{graphicx}
\begin{document}
\title{Quasinormal behavior of the D-dimensional Schwarzshild black hole and
higher order WKB approach}
\author{R.A.Konoplya \\
Department of Physics, Dniepropetrovsk National University\\
St. Naukova 13, Dniepropetrovsk  49050, Ukraine\\
konoplya@ff.dsu.dp.ua}
\date{}
\maketitle
\thispagestyle{empty}
\begin{abstract}
We study characteristic (quasinormal) modes of a $D$-dimensional
Schwarzshild black hole. It proves out that the real parts of the complex
quasinormal modes, representing the real oscillation frequencies,
are proportional to the product of the number of dimensions and inverse horizon radius
$\sim$ $D$  $r_{0}^{-1}$.  The asymptotic formula for large multipole number $l$ and
arbitrary $D$ is derived. In addition the WKB formula for computing QN modes,
developed to the 3th order beyond the eikonal approximation, is extended to the
6th order here. This gives us an accurate and economic way to compute
quasinormal frequencies.

\end{abstract}

\newpage

\section{Introduction}

Within the framework of the brane world models the size of  extra spatial dimensions may
be much larger than the Plank's length, and the fundamental quantum gravity scale may be very low
($\sim$ Tev). When considering models with large extra dimensions the
black hole mass may be of order Tev., i.e. much smaller than the
Plank's mass. There is a possibility of production of such mini black
holes in particle collisions in colliders and in cosmic ray
experiments \cite{S.Dimopoulos}. Estimations show that these higher
dimensional black holes can be described by classical solutions of
vacuum Einstein equations. Thus the investigation of general
properties of these black holes, including perturbations and decay of different
fields around them, attracts considerable interest now (see for example \cite{Frolov1},
\cite{Cardoso-Lemos2} and references therein).

It is well-known that when perturbing black hole it undergoes damping oscillations
which are characterized by some complex eigenvalues of the wave equations
called {\it quasinormal frequencies}. Their
real parts represent the oscillation frequencies, while the imaginary
ones determine the damping rates of the modes. The quasinormal modes
(QN) of black holes (BH's) depend only on a black hole parameters and
not on a way in which they were excited. QN's are called, therefore,
''footprints" of a black hole.
Being a useful characteristic of black hole's dynamics, quasinormal modes are
studied also
within different contexts now: in Anti-de-Sitter/Conformal Field Theory (AdS/CFT)
correspondence (see for example \cite{Horowitz1}- \cite{Moss-Norman} and references therein),
because
of the possibility of observing quasinormal ringing of astrophysical
BH's (see \cite{Kokkotas-Schmidt} for a review), when considering thermodynamic properties
of black holes in loop quantum gravity \cite{Dreyer}-\cite{Motl}, in the context of
possible connection with critical collapse \cite{Horowitz1}, \cite{Konoplya3},
\cite{Kim}, \cite{KonoplyaPLB1}.

Thus it would be interesting to know, from different grounds, what happen
with QN spectrum a black hole
living in $D$ -dimensional space-time \cite{Motl1}, \cite{Cardoso-Lemos2}.
The present paper is two-fold: First we
extend the WKB method of Schutz, Will and Iyer for computing QN modes
from the 3th to the 6th order beyond the eikonal approximation (see
Sec.II and Appendix I). In a lot of physical situations this allows
us to compute the QNMs  accurately and quickly without resorting to
complicated numerical methods. In Appendix II QN modes of $D=4$
Schwarzshild black hole induced by perturbations of different spin
are obtained by the 6th order WKB formula, and compared with the
numerical values and 3th order WKB values. Second, motivated by the
above reasons, we apply the obtained WKB formula for finding of the scalar quasinormal modes of
multi-dimensional Schwarzshild black hole (Sec. III). It proves out that the
real parts of the quasinormal frequencies are proportional to the
product $D r_{0}^{-1}$, where $r_{0}$ is the horizon radius, and $D$ is the
dimension of space-time.

\section{Sixth order WKB analysis}

First semi-analytical method for calculations of BH QNMs was apparently
proposed by Bahram Mashhoon who used the Poschl-Teller potential to estimate the
QN frequencies \cite{Mashhoon}.
In \cite{Will-Schutz} there was proposed a semi-analytical method for computing QNM's
based on the WKB treatment. Then in \cite{S.Iyer-C.M.Will}
the first WKB order formula  was extended to the third order
beyond the eikonal approximation, and, afterwards, was frequently used in a lot of works
(see for example \cite{Konoplya3} \cite{Kokkotas2}, \cite{Kokkotas3},  \cite{Simone-Will},  \cite{Andersson},
\cite{Piazza},  \cite{Onozawa}, \cite{KonoplyaGRG1} and references therein).
The accuracy of the 3th order WKB
formula (see eq. (1.5) in \cite{S.Iyer-C.M.Will}) is the better, the more multipole number $l$
and the less overtone $n$. For the Schwarzshild BH  the results practically
coincide with accurate numerical results of Leaver  \cite{Leaver} at $l\geq 4$ when being
restricted by lower overtones for which $l > n$. For fewer multipoles, however, accuracy
is worse, and may reaches $10$ per cents at $l=0$, $n=0$.
Numerical approach \cite{Leaver}, on contrary, is very accurate, but,
dealing with numerical integration or systems of recurrence
relations, is very cumbersome, and, often, require modification to be
applied to different effective potentials. At the same time WKB approach
lets us to obtain QNM's for a full range of parameters giving thereby
some fields of work for intuition as to physical behavior of a
system. Even though WKB formula gives the best accuracy at $l > n$,
it includes the case of astrophysical black hole radiation where only lower
overtones are significantly excited \cite{Stark}.
Both advantages and deficiencies of the WKB approach motivated us to
extend the existent 3th order WKB formula up to the 6th order.

The perturbation equations of a black hole can be reduced to the
Schrodinger wave-like equation:
\begin{equation}\label{1}
\frac{d^2 \psi}{d x^2} + Q(x) \psi(x)=0,
\end{equation}
where "the potential" $-Q(x)$ is constant at the event horizon
($x=-\infty$) and at the infinity ($x=+\infty$) and it rises to
maximum at some intermediate $x= x_{0}$. Consider radiation of a
given frequency $\omega$ incident on the black hole from infinity and
let $R(\omega)$ and $T(\omega)$ be the reflection and transmission
amplitudes respectively. Extend $R(\omega)$ to the complex frequency
plane such that $Re(z)\neq 0$, and $T(z)/R(z)$ is regular. Then, the
quasinormal modes correspond to the singularities of $R(z)$.
We have a direct analogy with the problem of scattering near the pick
of the potential barrier in quantum mechanics, where $\omega^2$ plays
a role of energy, and the two turning points divide the space into three
regions at which boundaries the corresponding solutions should be
matched.

To extend the 3th order WKB formula of \cite{S.Iyer-C.M.Will} we used
the technique of Iyer and Will. We shall omit here the
technicalities of this approach which are described in
\cite{S.Iyer-C.M.Will}.
The only thing we should stress is that since the coefficients
$M_{ij}$,
that connect amplitudes near the horizon with those at infinity,
depend only on $\nu$ (related to the overtone number $n$) they may be found to higher orders, simply by
solving the interior (between the turning points) problem to higher
orders. Thus there is no need to perform an explicit match of the solutions to
WKB solutions in the exterior (outside turning points) regions to the
same order. The result has the form:
\begin{equation}\label{2}
\frac{\imath Q_{0}}{\sqrt{2 Q_{0}''}}
-\Lambda_{2}-\Lambda_{3} -\Lambda_{4} -\Lambda_{5} -\Lambda_{6} =n+\frac{1}{2},
\end{equation}
where the correction terms $\Lambda_{4}$,  $\Lambda_{5}$,
$\Lambda_{6}$ can be found in the Appendix I. Note that  $\Lambda_{4}$ coincides with
preliminary formula (A3) of \cite{S.Iyer-C.M.Will} in proper designations.

An alternative, pure algebraic approach to finding higher order WKB corrections was
proposed by O.Zaslavskii \cite{O.Zaslavsky}, using a quantum
anharmonic oscillator problem where WKB correction terms come from
perturbation theory corrections to the potential anharmonicity.

Thus we have obtained an economic and accurate formula for straightforward calculation of QNM frequencies.
The 6th order formula applied to the $D=4$ Schwarzshild BH is as
accurate already at $l=1$ as the 3th order formula does at $l=4$.
We show it in Appendix II on example of QNM's
corresponding to perturbations of fields of different spins:
scalar ($s=0$), neutrino ($s=1/2$), electromagnetic ($s=1$),
gravitino ($s=3/2$), and gravitational ($s=2$).
In addition, looking at the convergence of all sixth WKB values to
some unknown true QN mode, we can judge, approximately, how far from the true QN
value we are, staying within the framework of WKB method.

\section{Quasinormal modes of the D-dimensional \\
Schwarzshild black hole}

The metric of the Schwarzshild black hole in $D$-dimensions has the
form:
\begin{equation}\label{3}
ds^2= f(f) dt^2 -f^{-1}(r) dr^2 + r^2 d \Omega^{2}_{D-2},
\end{equation}
where
\begin{equation}\label{4}
f(r)=1-\left(\frac{r_{0}}{r}\right)^{D-3}= 1-\frac{16 \pi G M}{(D-2) \Omega_{D-2} r^{D-3}}.
\end{equation}
Here we used the quantities
$$\Omega_{D-2}=\frac{(2 \pi)^{(D-1)/2}}{\Gamma((D-1)/2)}, \quad \Gamma(1/2) =
\sqrt{\pi}, \quad \Gamma(z+1)=z \Gamma(z)$$.

The scalar perturbation equation of this black hole can be reduced to the
Schrodinger wave-like equation (\ref{1}) with respect to the "tortoise" coordinate $x$:
$d x = \frac{d r}{f(r)}$
where "the potential" $-Q(x)$ has the form:
\begin{equation}\label{5}
Q(x)= \omega^2-f(r)\left(\frac{l(l+D-3)}{r^2} + \frac{(D-2) (D-4)}{4 r^2} f(r) + \frac{D-2}{2 r} f'(r)\right),
\end{equation}
At some fixed $D$ we can put $r_{0}=2$ and measure $\omega$ in units $2 r_{0}^{-1}$.
The quasinormal modes satisfy the boundary conditions:
\begin{equation}\label{6}
\phi(x) \sim c_{\pm} e^{ \pm i \omega x} \qquad as \qquad x\rightarrow \pm
\infty.
\end{equation}

The 6th WKB order formula used here gives very accurate results for low overtones.
The previous orders serve us to see the convergence of the WKB
values of $\omega^2$ as a WKB order grows to an accurate numerical result.
Namely we can observe that for $l=1,2,3,4,..$ for the fundamental overtone the
6th order values differs form its 5th order value by fractions of a percent or less at not
very large $D$ (we are restricted here by $D=4, 5, ...15$).

It proves out that if one takes $r_{0}=2$ for each given $D$, then
the real parts of $\omega$ for different $D$ lay on a strict
line. That is, $\omega_{Re}$ is proportional to the product $r_{0}
D$ (Remember that $r_{0}$ depends on $D$ itself).
Namely, for the fundamental overtone we obtain the
following approximate relations:
\begin{equation}\label{7}
\omega_{Re}  \sim  0.244 D (r_{0}/2)^{-1}, \qquad l=2
\end{equation}

\begin{equation}\label{8}
\omega_{Re}  \sim 0.275 D (r_{0}/2)^{-1}, \qquad l=3
\end{equation}

\begin{equation}\label{9}
\omega_{Re}  \sim 0.290 D (r_{0}/2)^{-1}, \qquad l=4.
\end{equation}
Here we take $\omega= \omega_{Re}- i \omega_{Im}$.
Generally, the more the multipole number $l$, the more the coefficient before the product $D r_{0}^{-1}$.
The same  $ \sim D r_{0}^{-1}$ relation we observed for higher overtone but not higher than $l$,
for which WKB treatment is applicable. In Fig.1,2 we presented the real and imaginary parts
of $\omega$ measured in $2 r_{0}^{-1}$ for different $D$. For real parts of $l=1$ modes we see the
deviation from the strict line at large $D$. This however,
is stipulated by a bad accuracy of the WKB approach, and we believe that the true frequencies will lay on
strict line again. Indeed, one can judge
about it by looking at the convergence plot Fig.3-Fig.6 where the
real and imaginary parts of $\omega$ are shown as a function of the
WKB order. Generally the accuracy of the WKB formula is the better,
the more $l$, and the less $n$ and $D$.  Note that the
dependence $D r_{0}^{-1}$ for lower overtones can be recovered even within
3th order formula, provided $l$ is greater than $2$, and $D$ is not very large.

\begin{figure}
\begin{center}
\includegraphics{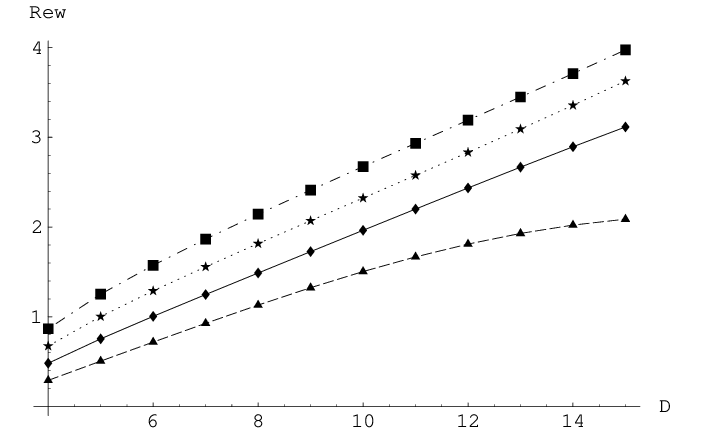}
\caption{$Re \omega$ for different dimensions $D$; $l=1$ (bottom), $2$, $3$, $4$ (top); $n=0$.}
\label{}
\end{center}
\end{figure}

\begin{figure}
\begin{center}
\includegraphics{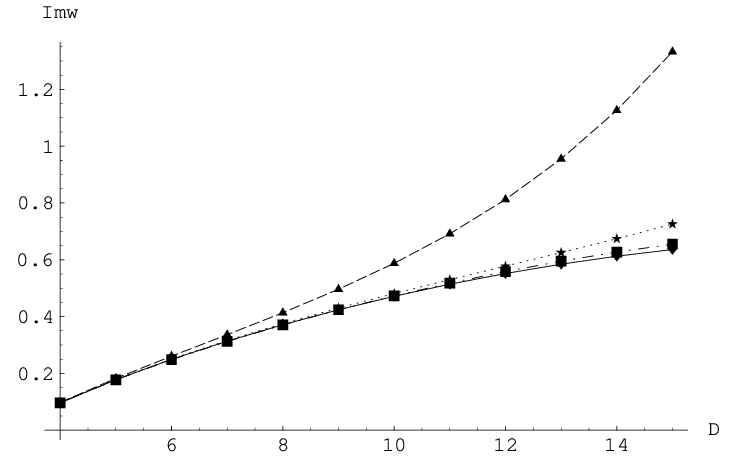}
\caption{$Im \omega$ for different dimensions $D$; $l=1$ (bottom), $2$, $3$, $4$ (top); $n=0$.}
\label{}
\end{center}
\end{figure}

\begin{figure}
\begin{center}
\includegraphics{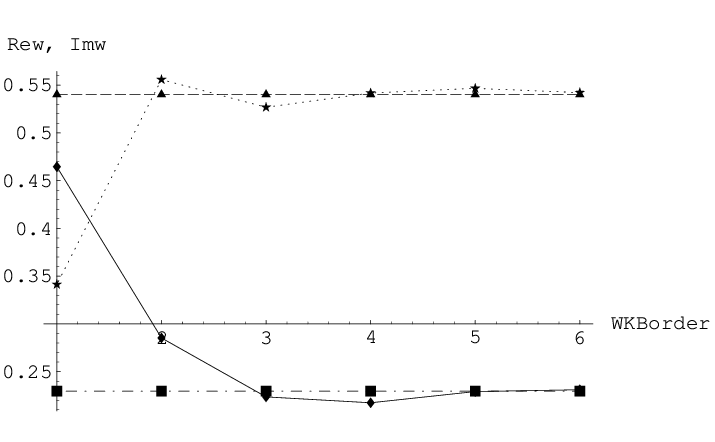}
\caption{$\omega_{Re}$ (bottom) and $\omega_{Im}$  (top) as a function of WKB order of
the formula with which it was obtained
for $l=1$, $n=2$, $D=4$ modes, and the corresponding
numerical value. We see how the WKB values converge to an accurate numerical value as the WKB order increases.}
\label{}
\end{center}
\end{figure}

\begin{figure}
\begin{center}
\includegraphics{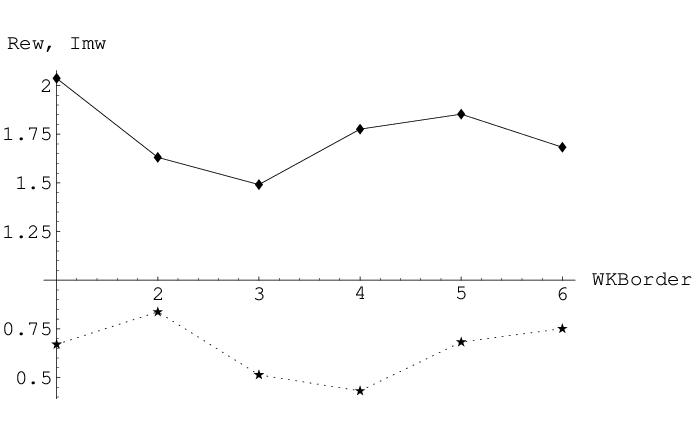}
\caption{$\omega_{Re}$ (top) and $\omega_{Im}$ (bottom) as a function of WKB order of the formula with
which it was obtained for $l=0$, $n=0$, $D=12$ modes.}
\label{}
\end{center}
\end{figure}

\begin{figure}
\begin{center}
\includegraphics{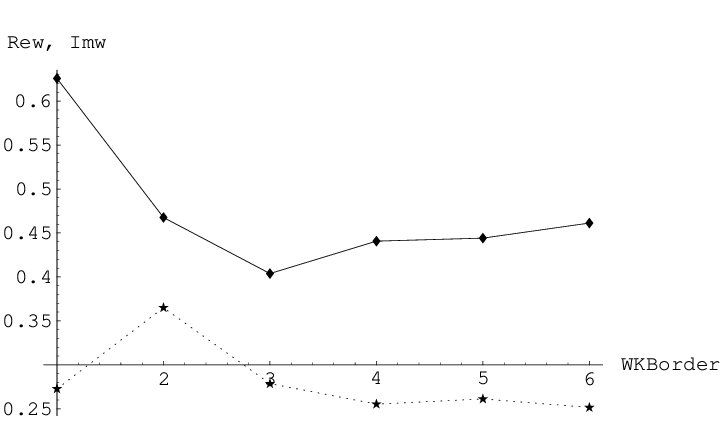}
\caption{$\omega_{Re}$ (top) and $\omega_{Im}$ (bottom) as a function of WKB order of the formula with which it was obtained
for $l=0$, $n=0$, $D=6$ modes.}
\label{}
\end{center}
\end{figure}

\begin{figure}
\begin{center}
\includegraphics{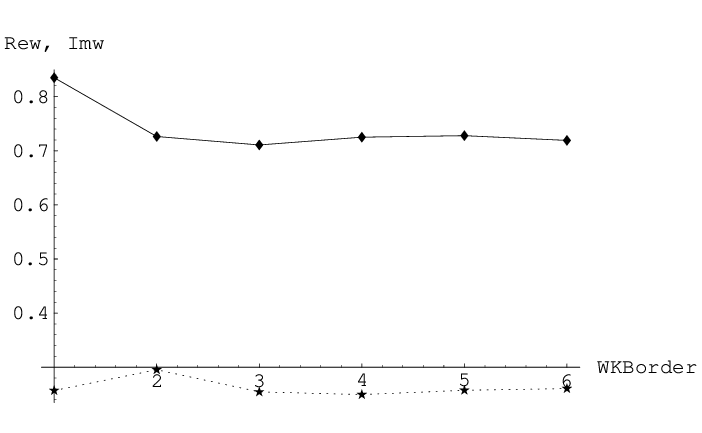}
\caption{$\omega_{Re}$ (top) and $\omega_{Im}$ (bottom) as a function of WKB order of the formula with which it was obtained
for $l=1$, $n=0$, $D=6$ modes.}
\label{}
\end{center}
\end{figure}

Another point is the $l=0$ modes: in this case the lowest overtone
implies $l=n$, and the WKB formula has considerable relative error.
For a four-dimensional BH, for which the accurate numerical results
are known, the error is about $10$ percent for $\omega_{Im}$, and $5$
percent for $\omega_{Re}$ in the third WKB order, while in the sixth
order it reduces to $0$ percent for $\omega_{Re}$ and $3$ percent for
$\omega_{Im}$ (see Appendix II). For greater $D$ the error increases,
the difference between the fifth and sixth order WKB values grows and
one cannot judge of true quasinormal behavior in this case (see Fig.3, 4, 5.).
Fortunately, other field perturbations, including gravitational, have
the lowest overtone with $l>n$ and the WKB treatment is of good accuracy for all $l$.
In Table 1. we compare the third order WKB values of $l=0$, $n=0$ modes for
different $D$  \cite{Cardoso-Lemos2} with those obtained through the sixth order here.

\begin{table}
\begin{center}
\begin{tabular}{|c|c|c|c|}
\hline
  $D$ & 3th WKB order & 6th WKB order & $1/r_{0}$ \\ \hline
  $4$ & $0.1046-0.1152 i$ & $0.1105 - 0.1008 i$ & $0.5$ \\
  $6$ & $1.0338- 0.7133 i$ & $1.1808- 0.6438 i$ & $1.28$\\
  $8$ & $1.9745- 1.0258 i$ & $2.3004- 1.0328 i$ & $1.32$\\
  $10$ & $2.7828- 1.1596 i$ & $3.2214- 1.3766 i$ & $1.25$\\
  $12$ & $3.4892- 1.2020 i$& $3.9384- 1.7574 i$ & $1.17$\\ \hline
\end{tabular}
\end{center}
Table I. Schwarzshild QN frequencies for $l=0$, $n=0$ scalar perturbations in various $D$.
\end{table}

For large $l$ the well-known approximate formula reads
(see \cite{Mashhoon2}, \cite{Will2}, \cite{Press} for a proof)
\begin{equation}\label{10}
\omega_{Re}= \frac{1}{3 \sqrt{3}}\left(l+\frac{1}{2}\right),
\qquad \omega_{Im}= \frac{1}{3 \sqrt{3}}\left(n+\frac{1}{2}\right)
\end{equation}

To obtain its $D$-dimensional generalization we find a value
$r_{max}$ at which the effective potential $V$ attains its maximum,
provided $l$ is large
\begin{equation}\label{11}
r_{max} \approx 2^{\frac{D-4}{D-3}} (D-1)^\frac{1}{D-3}, \qquad D=4,5,6,
\ldots .
\end{equation}
Then let us make use of this value $r_{max}$ when dealing with the first order WKB
formula. After expansion in terms of small values of $1/l$, for a fixed $D$
in units of $2 r_{0}^{-1}$ we obtain
\begin{equation}\label{12}
\omega_{Re} \approx \frac{D+2 l -3}{4} \left(\frac{2}{D-1}\right)^{\frac{1}{D-3}} \sqrt{\frac{D-3}{D-1}}
\end{equation}

\begin{equation}\label{13}
\omega_{Im} \approx \frac{(D-3)}{4} \left(\frac{2}{D-1}\right)^{\frac{1}{D-3}}
\frac{2 n +1}{\sqrt{D-1}}
\end{equation}

When $D=4$ these formulas go over into (\ref{11}). We see that when
$l$ is much larger than $D$, the $ \sim D r_{0}^{-1}$ dependence of
$\omega_{Re}$ breaks down.

\section{Conclusion}

We were interested here in a question how dimensionality effects on quasinormal
behavior of black holes.
Yet, several interesting points are left beyond our consideration of
low laying quasinormal modes of multi-dimensional black holes.
First of all, one would like to understand the origin of the relation
$ \sim D r_{0}^{-1}$ in $\omega_{Re}$ dependence. In this question it is
possible to try to explain it from the interpretations of QN modes as
Breit-Wigner type resonances generated by a family of surfaces wave
propagating close to the unstable circular photon orbit \cite{Decanini}.
Second, we do not know whether $ \sim D r_{0}^{-1}$ dependence will be
present for perturbations of other fields, and for more general
backgrounds, such as multi-dimensional Reissner-Nordstrom or Kerr.
We hope further investigations will clarify these points.

\section*{Acknowledgements}
It is a pleasure to acknowledge stimulating discussions with
Vitor Cardoso and Oleg Zaslavskii.


\section{Appendix 1: Correction terms for WKB formula}
Here we shall follow the designations: $Q_{0}$ means the value of the
potential $Q$ at its pick, while $Q_{i}$ is the $i$th derivative of
$Q$ with respect to the tortoise coordinate $x$. Then $Q_{i}^j$ is
the $j$th power of the $i$th derivative of $Q$.

$$\Lambda_{4}=\frac{1}{597196800 \sqrt{2} Q_{2}^{7} \sqrt{Q_{2}}}
 (2536975 Q_{3}^{6} - 9886275 Q_{2} Q_{3}^{4} Q_{4}
 +5319720 Q_{2}^{2} Q_{3}^{3} Q_{5}-$$
$$225 Q_{2}^{2} Q_{3}^{2}
 (-40261 Q_{4}^{2}+9688 Q_{2} Q_{6}) + 3240 Q_{2}^{3} Q_{3}
 (-1889 Q_{4} Q_{5}+220 Q_{2} Q_{7}) -$$
$$729 Q_{2}^{3} (1425 Q_{4}^{3}
 -1400 Q_{2} Q_{4} Q_{6} +8 Q_{2} (-123 Q_{5}^{2} +25 Q_{2}
 Q_{8})))+$$
 $$\frac{(n+1/2)^2}{4976640 \sqrt{2}
 Q_{2}^{7} \sqrt{Q_{2}}}
 (348425 Q_{3}^{6} -1199925 Q_{2} Q_{3}^{4} Q_{4}
 +57276 Q_{2}^{2} Q_{3}^{3} Q_{5}-$$
 $$45 Q_{2}^{2} Q_{3}^{2}
 (-20671 Q_{4}^{2}+4552 Q_{2} Q_{6}) + 1980 Q_{2}^{3} Q_{3}
 (-489 Q_{4} Q_{5}+52 Q_{2} Q_{7})-$$
 $$27 Q_{2}^{3} (2845 Q_{4}^{3}
 -2360 Q_{2} Q_{4} Q_{6} +56 Q_{2} (-31 Q_{5}^{2} +5 Q_{2}
 Q_{8})))+$$
 $$\frac{ (n+1/2)^4}{2488320 \sqrt{2} Q_{2}^{7}
 \sqrt{Q_{2}}}
 (192925 Q_{3}^{6} -581625 Q_{2} Q_{3}^{4} Q_{4}
 +234360 Q_{2}^{2} Q_{3}^{3} Q_{5}-$$
 $$45 Q_{2}^{2} Q_{3}^{2}
 (-8315 Q_{4}^{2}+1448 Q_{2} Q_{6}) + 1080 Q_{2}^{3} Q_{3}
 (-161 Q_{4} Q_{5}+12 Q_{2} Q_{7}) -$$
\begin{equation}\label{75}
27 Q_{2}^{3} (625 Q_{4}^{3}
 -440 Q_{2} Q_{4} Q_{6} +8 Q_{2} (-63 Q_{5}^{2} +5 Q_{2}
 Q_{8})))
\end{equation}
$$\Lambda_{5}=\frac{(n+1/2)}{57330892800 Q_{2}^{10}}
(2768256 Q_{10} Q_{2}^7 -1078694575 Q_{3}^8+5357454900
Q_{2}Q_{3}^6 Q_{4}-$$
$$2768587920 Q_{2}^2 Q_{3}^5 Q_{5} +90 Q_{2}^2 Q_{3}^4
(-88333625 Q_{4}^2+ 12760664 Q_{2} Q_{6})-$$
$$4320 Q_{2}^3 Q_{3}^3
(-1451425 Q_{4} Q_{5}+91928 Q_{2} Q_{7}) - 27 Q_{2}^4
(7628525 Q_{4}^4 -9382480  Q_{2} Q_{4}^2 Q_{6} +$$
$$64 Q_{2}^2 (19277 Q_{6}^2  + 37764 Q_{5} Q_{7}) +
576 Q_{2} Q_{4} (-21577 Q_{5}^2 +2505 Q_{2} Q_{8}))+$$
$$540 Q_{2}^3 Q_{3}^2 (6515475 Q_{4}^3 -3324792 Q_{2} Q_{4} Q_{6}
+16 Q_{2} (-126468 Q_{5}^2 +12679 Q_{2} Q_{8}))-$$
$$432 Q_{2}^4 Q_{3}
(5597075 Q_{4}^2 Q_{5}-854160 Q_{2} Q_{4} Q_{7} +8 Q_{2} (-145417 Q_{5}
Q_{6}
 + 6685 Q_{2} Q_{9})))+$$
$$\frac{(n+1/2)^3}{477757440 Q_{2}^{10}}
(31104 Q_{10} Q_{2}^7 -42944825 Q_{3}^8+193106700
Q_{2}Q_{3}^6 Q_{4}-$$
$$-90039120 Q_{2}^2 Q_{3}^5 Q_{5}+30 Q_{2}^2 Q_{3}^4
(-8476205 Q_{4}^2+1102568 Q_{2} Q_{6})-$$
$$4320 Q_{2}^3 Q_{3}^3
(-41165  Q_{4} Q_{5}+2312  Q_{2} Q_{7}) - 9 Q_{2}^4
(445825 Q_{4}^4-472880 Q_{2} Q_{4}^2 Q_{6} +$$
$$64 Q_{2}^2 (829 Q_{6}^2  + 1836 Q_{5} Q_{7}) +
4032 Q_{2} Q_{4} (-179 Q_{5}^2 +15 Q_{2} Q_{8}))+$$
$$180 Q_{2}^3 Q_{3}^2 (532615 Q_{4}^3 -241224 Q_{2} Q_{4} Q_{6}
+16 Q_{2} (-9352 Q_{5}^2 +799 Q_{2} Q_{8}))-$$
$$144 Q_{2}^4 Q_{3}
(392325 Q_{4}^2 Q_{5}-51600 Q_{2} Q_{4} Q_{7} +8 Q_{2} (-8853 Q_{5} Q_{6}
 + 335 Q_{2} Q_{9})))+$$
$$\frac{(n+1/2)^5}{1194393600 Q_{2}^{10}}
(10368 Q_{10} Q_{2}^7 -66578225 Q_{3}^8+272124300
Q_{2}Q_{3}^6 Q_{4}-$$
$$112336560  Q_{2}^2 Q_{3}^5 Q_{5} +9450 Q_{2}^2 Q_{3}^4
(-33775 Q_{4}^2 + 3656 Q_{2} Q_{6})-$$
$$151200 Q_{2}^3 Q_{3}^3
(-1297 Q_{4} Q_{5}+56 Q_{2} Q_{7}) - 27 Q_{2}^4
(89075 Q_{4}^4 -83440 Q_{2} Q_{4}^2 Q_{6} +$$
$$64 Q_{2}^2 (131 Q_{6}^2  + 396 Q_{5} Q_{7}) +
576 Q_{2} Q_{4} (-343 Q_{5}^2 +15 Q_{2} Q_{8}))+$$
$$540 Q_{2}^3 Q_{3}^2 (188125 Q_{4}^3 -71400 Q_{2} Q_{4} Q_{6}
+16 Q_{2} (-3052 Q_{5}^2 +177 Q_{2} Q_{8}))-$$
\begin{equation}\label{76}
432 Q_{2}^4 Q_{3}
(118825 Q_{4}^2 Q_{5}-11760 Q_{2} Q_{4} Q_{7} +8 Q_{2}
(-2303 Q_{5} Q_{6} + 55 Q_{2} Q_{9})))
\end{equation}
$$\Lambda_{6}=\frac{-i}{202263389798400 Q_{2}^{12}\sqrt{2 Q_{2}}}
(-171460800 Q_{12} Q_{2}^9 +1714608000
Q_{11} Q_{2}^8 Q_{3}-$$
$$10268596800 Q_{10} Q_{2}^7 Q_{3}^2 +
970010662775 Q_{3}^{10} + 3772137600 Q_{10} Q_{2}^8 Q_{4}-$$
$$6262634175525 Q_{2} Q_{3}^8 Q_{4}
+13782983196150 Q_{2}^{2} Q_{3}^{6} Q_{4}^{2} -11954148125850 Q_{2}^3
Q_{3}^4 Q_{4}^3 + $$
$$3449170577475 Q_{2}^4 Q_{3}^2 Q_{4}^4 -144528059025 Q_{2}^5 Q_{4}^5
+ 3352602187200 Q_{2}^2 Q_{3}^7 Q_{5}-$$
$$12300730092000 Q_{2}^3 Q_{3}^5 Q_{4} Q_{5} + 11994129604800
Q_{2}^4 Q_{3}^3 Q_{4}^2 Q_{5} -2624788605600
Q_{2}^5 Q_{3} Q_{4}^3 Q_{5}+$$
$$2580769643760  Q_{2}^4 Q_{3}^4 Q_{5}^2 -3453909784416
Q_{2}^5 Q_{3}^2 Q_{4} Q_{5}^2 +438440697072  Q_{2}^6 Q_{4}^2 Q_{5}^2+$$
$$+260524397952 Q_{2}^6 Q_{3} Q_{5}^3-1475306441280 Q_{2}^3 Q_{3}^6 Q_{6}
+4329682610400 Q_{2}^4 Q_{3}^4 Q_{4} Q_{6}-$$
$$2865128172480 Q_{2}^5 Q_{3}^2 Q_{4}^2 Q_{6} +
233443879200 Q_{2}^6 Q_{4}^3 Q_{6} -1660199804928 Q_{2}^5 Q_{3}^3 Q_{5} Q_{6}+$$
$$1281705296256 Q_{2}^6 Q_{3} Q_{4} Q_{5} Q_{6} -87403857408  Q_{2}^7 Q_{5}^2 Q_{6} + 231105873600
Q_{2}^6 Q_{3}^2 Q_{6}^2-$$
$$ 68412859200 Q_{2}^7 Q_{4} Q_{6}^2  +552968700480 Q_{2}^4 Q_{3}^5 Q_{7}
-1231789749120 Q_{2}^5 Q_{3}^3 Q_{4} Q_{7}+$$
$$470726303040 Q_{2}^6 Q_{3} Q_{4}^2 Q_{7}   +413953400448 Q_{2}^6 Q_{3}^2 Q_{5} Q_{7}
-126242178048 Q_{2}^7 Q_{4} Q_{5} Q_{7}-$$
$$ 91489305600Q_{2}^7 Q_{3} Q_{6} Q_{7} + 5619715200 Q_{2}^8 Q_{7}^2
-175752294480 Q_{2}^5 Q_{3}^4 Q_{8}+$$
$$ 271759652640  Q_{2}^6 Q_{3}^2 Q_{4} Q_{8} -39736040400 Q_{2}^7 Q_{4}^2 Q_{8}-
73378363968 Q_{2}^7 Q_{3} Q_{5} Q_{8}+$$
$$9773265600  Q_{2}^8 Q_{6} Q_{8} +47107126080  Q_{2}^6 Q_{3}^3 Q_{9}
-43345290240  Q_{2}^7 Q_{3} Q_{4} Q_{9} +7400248128 Q_{2}^8 Q_{5} Q_{9})-$$

$$\frac{(n+1/2)^2 i}{687970713600 Q_{2}^{12}\sqrt{2 Q_{2}}}
(-4551552 Q_{12} Q_{2}^9+60279552
Q_{11} Q_{2}^8 Q_{3}-$$
$$425036160 Q_{10} Q_{2}^7 Q_{3}^2 +
73727194625 Q_{3}^{10} +116743680 Q_{10} Q_{2}^8 Q_{4}-$$
$$443649208275 Q_{2} Q_{3}^8 Q_{4}+
901144103850 Q_{2}^{2} Q_{3}^{6} Q_{4}^{2} -711096726150 Q_{2}^3
Q_{3}^4 Q_{4}^3 + $$
$$182164306725 Q_{2}^4 Q_{3}^2 Q_{4}^4 -6289615575 Q_{2}^5 Q_{4}^5
+ 222467624400 Q_{2}^2 Q_{3}^7 Q_{5}-$$
$$746418445200 Q_{2}^3 Q_{3}^5 Q_{4} Q_{5} + 653423900400
Q_{2}^4 Q_{3}^3 Q_{4}^2 Q_{5} -124319674800
Q_{2}^5 Q_{3} Q_{4}^3 Q_{5}+$$
$$143980943040 Q_{2}^4 Q_{3}^4 Q_{5}^2 -169712521920
Q_{2}^5 Q_{3}^2 Q_{4} Q_{5}^2 +18188188416 Q_{2}^6 Q_{4}^2 Q_{5}^2+$$
$$11240861184 Q_{2}^6 Q_{3} Q_{5}^3-91198200240 Q_{2}^3 Q_{3}^6 Q_{6}
+241513732080 Q_{2}^4 Q_{3}^4 Q_{4} Q_{6}-$$
$$140030897040 Q_{2}^5 Q_{3}^2 Q_{4}^2 Q_{6}
+9200103120 Q_{2}^6 Q_{4}^3 Q_{6} -84218693760 Q_{2}^5 Q_{3}^3 Q_{5} Q_{6}+$$
$$55248386688 Q_{2}^6 Q_{3} Q_{4} Q_{5} Q_{6} -3173043456 Q_{2}^7 Q_{5}^2 Q_{6} + 10464952896
Q_{2}^6 Q_{3}^2 Q_{6}^2-$$
$$ 2403421632 Q_{2}^7 Q_{4} Q_{6}^2  +31637744640 Q_{2}^4 Q_{3}^5 Q_{7}
-62649953280 Q_{2}^5 Q_{3}^3 Q_{4} Q_{7}+$$
$$20409822720 Q_{2}^6 Q_{3} Q_{4}^2 Q_{7}   + 18860532480 Q_{2}^6 Q_{3}^2 Q_{5} Q_{7}
-4693344768 Q_{2}^7 Q_{4} Q_{5} Q_{7}-$$
$$  3625731072 Q_{2}^7 Q_{3} Q_{6} Q_{7} + 188054784 Q_{2}^8 Q_{7}^2
-9155635200 Q_{2}^5 Q_{3}^4 Q_{8}+$$
$$ 12238024320 Q_{2}^6 Q_{3}^2 Q_{4} Q_{8} -1405278720 Q_{2}^7 Q_{4}^2 Q_{8}-
2866700160 Q_{2}^7 Q_{3} Q_{5} Q_{8}+$$
$$303295104 Q_{2}^8 Q_{6} Q_{8} +2210705280 Q_{2}^6 Q_{3}^3 Q_{9}
-1685525760  Q_{2}^7 Q_{3} Q_{4} Q_{9} +235488384 Q_{2}^8 Q_{5} Q_{9})-$$

$$\frac{(n+1/2)^4 i}{20065812480 Q_{2}^{12}\sqrt{2 Q_{2}}}
(-66528 Q_{12} Q_{2}^9+1245888
Q_{11} Q_{2}^8 Q_{3}-$$
$$11158560 Q_{10} Q_{2}^7 Q_{3}^2 +
4668804525 Q_{3}^{10} +2116800 Q_{10} Q_{2}^8 Q_{4}-$$
$$25898331375 Q_{2} Q_{3}^8 Q_{4}+
47959232650 Q_{2}^{2} Q_{3}^{6} Q_{4}^{2} -33861927750 Q_{2}^3
Q_{3}^4 Q_{4}^3 + $$
$$7454763225 Q_{2}^4 Q_{3}^2 Q_{4}^4 -184988475 Q_{2}^5 Q_{4}^5
+ 11891917800 Q_{2}^2 Q_{3}^7 Q_{5}-$$
$$36105463800 Q_{2}^3 Q_{3}^5 Q_{4} Q_{5} + 27953667000
Q_{2}^4 Q_{3}^3 Q_{4}^2 Q_{5} -4457716200
Q_{2}^5 Q_{3} Q_{4}^3 Q_{5}+$$
$$6285855240 Q_{2}^4 Q_{3}^4 Q_{5}^2 -6471756144
Q_{2}^5 Q_{3}^2 Q_{4} Q_{5}^2 +565259688 Q_{2}^6 Q_{4}^2 Q_{5}^2+$$
$$380939328 Q_{2}^6 Q_{3} Q_{5}^3-4375251160 Q_{2}^3 Q_{3}^6 Q_{6}
+10317018600 Q_{2}^4 Q_{3}^4 Q_{4} Q_{6}-$$
$$5113813320 Q_{2}^5 Q_{3}^2 Q_{4}^2 Q_{6}
+238888440 Q_{2}^6 Q_{4}^3 Q_{6} -3203871552 Q_{2}^5 Q_{3}^3 Q_{5} Q_{6}+$$
$$1758685824 Q_{2}^6 Q_{3} Q_{4} Q_{5} Q_{6} -88566912 Q_{2}^7 Q_{5}^2 Q_{6} + 335466432
Q_{2}^6 Q_{3}^2 Q_{6}^2-$$
$$ 55073088 Q_{2}^7 Q_{4} Q_{6}^2  +1351294560 Q_{2}^4 Q_{3}^5 Q_{7}
-2341442880  Q_{2}^5 Q_{3}^3 Q_{4} Q_{7}+$$
$$626542560 Q_{2}^6 Q_{3} Q_{4}^2 Q_{7} + 619520832 Q_{2}^6 Q_{3}^2 Q_{5} Q_{7}
-123524352 Q_{2}^7 Q_{4} Q_{5} Q_{7}-$$
$$96574464 Q_{2}^7 Q_{3} Q_{6} Q_{7} + 4048704 Q_{2}^8 Q_{7}^2
-341160120  Q_{2}^5 Q_{3}^4 Q_{8}+$$
$$386210160 Q_{2}^6 Q_{3}^2 Q_{4} Q_{8} -30837240 Q_{2}^7 Q_{4}^2 Q_{8}-
78073632 Q_{2}^7 Q_{3} Q_{5} Q_{8}+$$
$$5848416 Q_{2}^8 Q_{6} Q_{8} +70415520 Q_{2}^6 Q_{3}^3 Q_{9}
-43424640 Q_{2}^7 Q_{3} Q_{4} Q_{9} +5255712 Q_{2}^8 Q_{5} Q_{9})-$$

$$\frac{(n+1/2)^6 i}{300987187200Q_{2}^{12}\sqrt{2 Q_{2}}}
(-72576 Q_{12} Q_{2}^9+1886976
Q_{11} Q_{2}^8 Q_{3}-$$
$$22135680 Q_{10} Q_{2}^7 Q_{3}^2 +
27463538375 Q_{3}^{10} +2903040 Q_{10} Q_{2}^8 Q_{4}-$$
$$141448688325 Q_{2} Q_{3}^8 Q_{4}+
240655765350 Q_{2}^{2} Q_{3}^{6} Q_{4}^{2} -152907158250 Q_{2}^3
Q_{3}^4 Q_{4}^3 + $$
$$28724479875 Q_{2}^4 Q_{3}^2 Q_{4}^4 -413669025 Q_{2}^5 Q_{4}^5
+59058073200 Q_{2}^2 Q_{3}^7 Q_{5}-$$
$$164264209200  Q_{2}^3 Q_{3}^5 Q_{4} Q_{5} +113654696400
Q_{2}^4 Q_{3}^3 Q_{4}^2 Q_{5} -15166342800
Q_{2}^5 Q_{3} Q_{4}^3 Q_{5}+$$
$$26061194880 Q_{2}^4 Q_{3}^4 Q_{5}^2 -23876233920
Q_{2}^5 Q_{3}^2 Q_{4} Q_{5}^2 +1767189312 Q_{2}^6 Q_{4}^2 Q_{5}^2+$$
$$1292433408 Q_{2}^6 Q_{3} Q_{5}^3-18902165520 Q_{2}^3 Q_{3}^6 Q_{6}
+40256773200 Q_{2}^4 Q_{3}^4 Q_{4} Q_{6}-$$
$$17116974000 Q_{2}^5 Q_{3}^2 Q_{4}^2 Q_{6}
+483582960 Q_{2}^6 Q_{4}^3 Q_{6} -11384150400 Q_{2}^5 Q_{3}^3 Q_{5} Q_{6}+$$
$$5285056896 Q_{2}^6 Q_{3} Q_{4} Q_{5} Q_{6} -246903552 Q_{2}^7 Q_{5}^2 Q_{6} + 992779200
Q_{2}^6 Q_{3}^2 Q_{6}^2-$$
$$101860416 Q_{2}^7 Q_{4} Q_{6}^2  +4966859520 Q_{2}^4 Q_{3}^5 Q_{7}
-7661606400 Q_{2}^5 Q_{3}^3 Q_{4} Q_{7}+$$
$$1683037440 Q_{2}^6 Q_{3} Q_{4}^2 Q_{7} +1861574400 Q_{2}^6 Q_{3}^2 Q_{5} Q_{7}
-316141056 Q_{2}^7 Q_{4} Q_{5} Q_{7}-$$
$$235146240 Q_{2}^7 Q_{3} Q_{6} Q_{7} +8895744 Q_{2}^8 Q_{7}^2
-1042372800 Q_{2}^5 Q_{3}^4 Q_{8}+$$
$$1016789760 Q_{2}^6 Q_{3}^2 Q_{4} Q_{8} -52436160 Q_{2}^7 Q_{4}^2 Q_{8}-
189060480 Q_{2}^7 Q_{3} Q_{5} Q_{8}+$$
\begin{equation}\label{77}
9217152 Q_{2}^8 Q_{6} Q_{8} +175190400 Q_{2}^6 Q_{3}^3 Q_{9}
-87816960 Q_{2}^7 Q_{3} Q_{4} Q_{9} +10378368 Q_{2}^8 Q_{5} Q_{9})
\end{equation}

All six WKB corrections printed in MATEMATICA are available from the
author in electronic form upon request.

\section{Appendix 2: QNMs of a 4-dimensional Schwarzshild black hole}

"The potential" $Q(x)$ in case of a Schwarzshild black hole has the
form
\begin{equation}\label{3}
Q(x)= \omega^2-\left(1-\frac{1}{r}\right)\left(\frac{l(l+1)}{r^2}
+\frac{1-s^2}{r^3}\right),
\end{equation}
where $s=0$ corresponds to scalar perturbations, $s=1/2$ - neutrino
perturbations, $s=1$ - electromagnetic perturbations, $s=3/2$ -
gravitino perturbations, $s=2$ - gravitational perturbations. The
quasinormal frequencies at 3th and 6th WKB orders and in comparison
with numerical results \cite{Leaver} are presented in the
table I.

\begin{table}
\begin{center}
\begin{tabular}{|c|c|c|c|}
  \hline
  $s=0$ & numerical & 3th order WKB & 6th order WKB \\
  \hline
  $l=0$, $n=0$ & $0.1105- 0.1049 i$ & $0.1046-0.1152 i$ & $0.1105 - 0.1008 i$ \\
  $l=1$, $n=0$ & $0.2929- 0.0977 i$ & $0.2911- 0.0980 i$ & $0.2929- 0.0978 i$ \\
  $l=1$, $n=1$ & $0.2645- 0.3063 i$ & $0.2622- 0.3074 i$ & $0.2645- 0.3065 i$ \\
  $l=2$, $n=0$ & $0.4836 -0.0968 i$ & $0.4832 -0.0968 i$ & $0.4836 -0.0968 i$ \\
  $l=2$, $n=1$ & $0.4639 -0.2956 i$ & $0.4632 -0.2958 i$ & $0.4638 -0.2956 i$ \\
  $l=2$, $n=2$ & $0.4305 -0.5086 i$ & $0.4317 -0.5034 i$ & $0.4304 -0.5087 i$ \\ \hline
  \hline
  $s=1/2$ & numerical & 3th order WKB & 6th order WKB \\
  \hline
  $l=1$, $n=0$ & $-$ & $0.2803 - 0.0969 i$ & $0.2822 - 0.0967 i$ \\
  $l=1$, $n=1$ & $-$ & $0.2500 - 0.3049 i$ & $0.2525 - 0.3040 i$ \\
  $l=2$, $n=0$ & $-$ & $0.4768 - 0.9639 i$ & $0.4772- 0.0963 i$ \\
  $l=2$, $n=1$ & $-$ & $0.4565 - 0.2947 i$ & $0.4571 -0.2945 i$ \\
  $l=2$, $n=2$ & $-$ & $0.4244 - 0.5016 i$ & $0.4231 -0.5070 i$ \\
  $l=3$, $n=0$ & $-$ & $0.6706 - 0.0963 i$ & $0.6708 - 0.0963 i$ \\
  $l=3$, $n=1$ & $-$ & $0.6557 - 0.2917 i$ & $0.6560 - 0.2917 i$ \\
  $l=3$, $n=2$ & $-$ & $0.6299 - 0.4931 i$ & $0.6286 - 0.4950 i$ \\
  $l=3$, $n=3$ & $-$ & $0.5970 - 0.6997 i$ & $0.5932 - 0.7102 i$ \\\hline
  \hline
  $s=1$ & numerical & 3th order WKB & 6th order WKB \\
  \hline
  $l=1$, $n=0$ & $0.2483-0.0925 i$ & $0.2459-0.0931 i$ & $0.2482-0.0926 i$ \\
  $l=1$, $n=1$ & $0.2145-0.2937 i$ & $0.2113-0.2958 i$ & $0.2143-0.2941 i$ \\
  $l=2$, $n=0$ & $0.4576-0.0950 i$ & $0.4571-0.0951 i$ & $0.4576-0.0950 i$ \\
  $l=2$, $n=1$ & $0.4365-0.2907 i$ & $0.4358-0.2910 i$ & $0.4365-0.2907 i$ \\
  $l=2$, $n=2$ & $0.4012-0.5016 i$ & $0.4023-0.4959 i$ & $0.4009-0.5017 i$ \\
  $l=3$, $n=0$ & $0.6569-0.0956 i$ & $0.6567-0.0956 i$ & $0.6569-0.0956 i$ \\
  $l=3$, $n=1$ & $0.6417-0.2897 i$ & $0.6415-0.2898 i$ & $0.6417-0.2897 i$ \\
  $l=3$, $n=2$ & $0.6138-0.4921 i$ & $0.6151-0.4901 i$ & $0.6138-0.4921 i$ \\
  $l=3$, $n=3$ & $0.5779-0.7063 i$ & $0.5814-0.6955 i$ & $0.5775-0.7065 i$ \\\hline
   \hline
 $s=3/2$ & numerical & 3th order WKB & 6th order WKB \\
  \hline
  $l=1$, $n=0$ & $-$ & $0.1817 - 0.0866 i$ & $0.1739 - 0.08357 i$ \\
  $l=1$, $n=1$ & $-$ & $0.1354 - 0.2812 i$ & $0.1198 - 0.2813 i$ \\
  $l=2$, $n=0$ & $-$ & $0.4231 - 0.926 i$ & $0.4236- 0.0925 i$ \\
  $l=2$, $n=1$ & $-$ & $0.4000 - 0.2842 i$ & $0.4007 -0.2838 i$ \\
  $l=2$, $n=2$ & $-$ & $0.3636 - 0.4853 i$ & $0.3618 -0.4919 i$ \\
  $l=3$, $n=0$ & $-$ & $0.6332 - 0.0945 i$ & $0.6333 - 0.0944 i$ \\
  $l=3$, $n=1$ & $-$ & $0.6173 - 0.2864 i$ & $0.6175 - 0.2863 i$ \\
  $l=3$, $n=2$ & $-$ & $0.5898 - 0.4846 i$ & $0.5884 - 0.4868 i$ \\
  $l=3$, $n=3$ & $-$ & $0.5547 - 0.6882 i$ & $0.5505 - 0.7000 i$ \\\hline
    \hline
 $s=2$ & numerical & 3th order WKB & 6th order WKB \\
  \hline
  $l=2$, $n=0$ & $0.3737-0.0890 i$ & $0.3732-0.0892 i$ & $0.3736-0.0890 i$ \\
  $l=2$, $n=1$ & $0.3467-0.2739 i$ & $0.3460-0.2749 i$ & $0.3463-0.2735 i$ \\
  $l=2$, $n=2$ & $0.3011-0.4783 i$ & $0.3029-0.4711 i$ & $0.2985-0.4776 i$ \\
  $l=3$, $n=0$ & $0.5994-0.0927 i$ & $0.5993-0.0927 i$ & $0.5994-0.0927 i$ \\
  $l=3$, $n=1$ & $0.5826-0.2813 i$ & $0.5824-0.2814 i$ & $0.5826-0.2813 i$ \\
  $l=3$, $n=2$ & $0.5517-0.4791 i$ & $0.5532-0.4767 i$ & $0.5516-0.4790 i$ \\
  $l=3$, $n=3$ & $0.5120-0.6903 i$ & $0.5157-0.6774 i$ & $0.5111-0.6905 i$ \\
  $l=4$, $n=0$ & $0.8092-0.0942 i$ & $0.8091-0.0942 i$ & $0.8092-0.0942 i$ \\
  $l=4$, $n=1$ & $0.7966-0.2843 i$ & $0.7965-0.2844 i$ & $0.7966-0.2843 i$ \\
  $l=4$, $n=2$ & $0.7727-0.4799 i$ & $0.7736-0.4790 i$ & $0.7727-0.4799 i$ \\
  $l=4$, $n=3$ & $0.7398-0.6839 i$ & $0.7433-0.6783 i$ & $0.7397-0.6839 i$ \\
  $l=4$, $n=4$ & $0.7015-0.8982 i$ & $0.7072-0.8813 i$ & $0.7006-0.8985 i$ \\ \hline
\end{tabular}
\end{center}
Table I. Schwarzshild QN frequencies for perturbations of different spin.
\end{table}

\end{document}